\documentstyle[12pt]{article}
\begin{document}

\begin{flushright}
February 1995 \\
\end{flushright}

\vspace{2cm}

\begin{center}
{\Large Hadron Structure and  the QCD Phase Transition\footnote{
Invited talk at the 9-th Nishinomiya Yukawa-Memorial Symposium on
Theoretical Physics, (Oct. 27-28, 1994, Nishinomiya, Japan)}}\\
\vspace*{1cm}
{\large Tetsuo Hatsuda\footnote{e-mail address: hatsuda@nucl.ph.tsukuba.ac.jp}
 \\
        Institute of Physics, University of Tsukuba,\\
         Tsukuba, Ibaraki 305,  Japan}
\end{center}
\vspace*{1cm}

\begin{abstract}

  Firstly, I give a brief summary of the current
  understanding of QCD below and near $T_c$ (the critical temperature
 of the chiral transition). Some emphases are put on the
  qualitative difference between the Yukawa regime ($T \sim 0)$
  and the Hagedorn regime ($T \sim T_c$).
 Secondly, the dynamical phenomena associated with the chiral transition,
  in particular, the spectral changes
 of hadrons in hot and/or dense medium are reviewed from the
 point of view of the QCD spectral sum rules.
  Confusions on  the QCD sum rules at
 finite temperature/density  are also clarified and  remarks
 on the effective lagrangian approaches are given.
 Thirdly,  planned experiments to detect the  spectral changes
 in medium are summarized.
\end{abstract}

\newpage
\section*{Contents}

\noindent
1. Introduction

\vspace{0.3cm}

\noindent
 2. QCD phase transition and continuum percolation

\noindent
\ \ \ \ 2.1. Transition from Yukawa regime to Hagedorn regime

\noindent
\ \ \ \ 2.2. Continuum percolation and $T_c$

\vspace{0.3cm}

\noindent
 3. Phase transition on the lattice

\noindent
\ \ \ \ 3.1.  Energy density, pressure and chiral condensate

\noindent
\ \ \ \ 3.2.  Effective theory interpretation

\vspace{0.3cm}

\noindent
 4. Dynamical critical phenomena

\noindent
\ \ \ \ 4.1. Static fluctuation of the order parameter

\noindent
\ \ \ \ 4.2. Dynamical fluctuation and para-pion

\noindent
\ \ \ \ 4.3. Soft modes -- examples in solid state physics --

\vspace{0.3cm}

\noindent
 5. Vector mesons at finite temperature

\noindent
\ \ \ \ 5.1. Low $T$ theorem in the Yukawa regime

\noindent
\ \ \ \ 5.2. Approaching $T_c$ -- Hagedorn regime --

\noindent
\ \ \ \ 5.3. QCD sum rules at finite $T$

\noindent
\ \ \ \ 5.4. Spectral change in the Yukawa regime and Hagedorn regime

\noindent
\ \ \ \ 5.5. $\phi$-meson near $T_c$

\noindent
\ \ \ \ 5.6. Collision widths of vector mesons

\noindent
\ \ \ \ 5.7. $T$-dependent Wilson coefficients -- Are they meaningful? --

\noindent
\ \ \ \ 5.8. Remarks on  effective theory approaches

\vspace{0.3cm}

\noindent
 6. Partial restoration of chiral symmetry at finite density

\noindent
\ \ \ \ 6.1. Vector mesons in nuclear matter

\noindent
\ \ \ \  6.2. Constraints from QCD sum rules

\noindent
\ \ \ \ 6.3. Use and misuse of the QCD sum rules in nuclear matter

\noindent
\ \ \ \  6.4. Remarks on effective theory approaches

\vspace{0.3cm}

\noindent
7. Detection of the spectral changes in laboratories

\noindent
\ \ \ \ 7.1. Finite temperature case

\noindent
\ \ \ \  7.2. Finite density case

\vspace{0.3cm}

\noindent
 8. Concluding remarks

\newpage

\section{Introduction}

 Due to the asymptotic freedom, the QCD coupling constant
 $g(\mu)$ decreases logarithmically as one increases the renormalization
 scale $\mu$,
 \begin{eqnarray}
{g^2(\mu) \over {4\pi}} \rightarrow {{12\pi}
 \over {(33-2N_f)\ln(\mu^2/\Lambda^2)}}
\ \ \ .
 \end{eqnarray}
  $\mu$ is   chosen to be a typical scale of the system
 to suppress the higher order terms in $g$.  Thus,
 in extremely hot and/or dense QCD medium,
 $\mu$ should be proportional to $T$ (temperature) or the
 chemical potential,
 which means that  systems at high $T$ and/or high baryon density $\rho$
 are composed of  weakly interacting  quarks and gluons
   (quark-gluon plasma phase) \cite{CP}.
 At low $T-\rho$ on the contrary,
  quarks and gluons are confined inside mesons and baryons
  (hadronic phase). Therefore  one may expect
 a phase transition between the two phases at intermediate  $T$ and $\rho$.
 The existence of such transition at finite $T$
  is in fact ``proved'' by the numerical simulations
 of QCD formulated on the lattice  \cite{LAT}.
 RHIC at BNL and LHC at CERN will serve as  machines
 to create and detect the hot quark-gluon plasma
 through the  relativistic heavy-ion collisions \cite{Hayano}.
 Several experiments are also  planned to detect the
 partial restoration of chiral symmetry in  heavy nuclei
 through the reactions such as $\gamma+A \rightarrow A^* + e^+ e^-$ and
$p+A \rightarrow  A^* + e^+ e^- $ \cite{SE}.

 The fundamental questions here may be summarized as

\noindent
\ \ \ \ (i) What are the properties of the quark gluon plasma?

\noindent
\ \ \ \ (ii) What is the precise nature of the  phase transition?

\noindent
\ \ \ \ (iii) What sort of critical phenomena occurs ?

 In the following, I will discuss some of the recent topics related to the
above questions with main emphases on (iii).
 In Section 2, an intuitive view of the QCD phase transition
 based on the percolation picture is given.
 In Section 3, the natures of the deconfinement and chiral transition
 studied on the lattice are briefly summarized.
 In Section 4, general introduction to the
dynamical critical phenomena  associated with the
 chiral transition is given.
 In section 5, Spectral change of the vector mesons
 at finite $T$ is examined as a typical and observable example of the
 dynamical critical phenomena.
 In Section 6, possibility of the partial restoration of chiral symmetry
  in heavy nuclei and associated phenomena (such as the mass shift of
  the vector
 mesons) are discussed.  In Section 7, the planned experiments for
 detecting the spectral change of the vector mesons are summarized.
 Section 8 is devoted to  concluding remarks.

\section{QCD phase transition and continuum percolation}

\subsection{Transition from Yukawa regime to Hagedorn regime}

 Let us first start  with an intuitive picture of the QCD phase transition
  given in Fig.1.
 Imagine heating  up the  QCD vacuum.  At low $T$, pions (the lightest
 mode in QCD) are thermally excited. As one increases $T$,
  massive hadrons  are also excited.  Since the pions and other
 hadrons have their own
 size (e.g. the pion radius is  about $0.65$fm), the thermal hadrons
  start to overlap with each other
 at certain temperature $T_c$ and dissolve
 into a gas of quarks and gluons  above $T_c$.

\begin{figure}[t]
   \vspace{13pc}
   \caption{A schematic figure of the percolation transition
 in QCD at $T \neq 0$.}
\end{figure}

Now, what kind of hadrons are mainly populated near the critical temperature?
 To get a rough idea, let's look at  $n/T^3$ (number density of hadrons divided
 by $T^3$) as a function of $T$ given in Fig. 2.  The
  interactions between hadrons are neglected  there \cite{GL}.
  The dashed line in Fig. 2 is a contribution of hypothetical massless pions.
 $r$ denotes the sum of massive resonance contributions other than
 $\pi, K, \eta, \rho, \omega $.
 Although only  pions are excited below 100 MeV,
 the resonance contributions start to dominate over pions above 160 MeV.
 Why this happens?  The reason is that the state density $\rho(m)$ of the
 excited hadrons with mass  $m$
 increases exponentially as $m$ increases:
\begin{eqnarray}
n_{tot}(T) & = & \int_0^{\infty} dm \ n(m;T)\ \rho(m) , \nonumber \\
 & &  n(m;T)  \propto e^{-m/T}, \ \ \ \ \rho(m) = {C \over m^{\alpha} }
 e^{m/\tau} ,
\end{eqnarray}
where $n_{tot}(T)$ is the total number density of hadrons,
 $n(m;T)$ is the number density of a
resonance with mass $m$,
 and $\alpha = 2 \ (2.5)$, $\tau \simeq $ 200 MeV (140MeV)
  for statistical bootstrap model  (for string model).
 The Boltzman suppression factor $\exp (-m/T)$ in $n(m;T)$  is compensated
 by the exponentially growing factor $\exp (m/\tau)$ in $\rho(m)$ and
 one is not allowed to neglect the massive states
  any more at high $T$ \cite{HAG}.
 This is a unique feature of the confining theory like QCD.

\begin{figure}[t]
   \vspace{15pc}
\caption{ (Number density of hadrons)/$T^3$ as a function of $T$.
  $r$ denotes the sum of the resonance contributions
 other than  $\pi, K, \eta, \rho, \omega $ [5].}
 \end{figure}

 I will call the pion-dominated region ($T<$ 150 MeV) as {\bf
 Yukawa regime} and the resonance-dominated region ($T>$ 150 MeV)
  as {\bf Hegedorn regime} because of the obvious reason.
 Whether $T_c$ is in the Yukawa regime or in the Hegedorn regime
 is a dynamical question which is examined in the next subsection.

\subsection{Continuum percolation and $T_c$}

Fig. 1 together with Fig. 2 indicates that we may be able to use
 the knowledge of the so-called continuum or random percolation
 problem \cite{Perco} to estimate $T_c$.
 The continuum percolation problem is defined as follows.
 Let us prepare $N$ set of  spheres with the same radius $R$.
Then distribute them randomly in a three dimensional box with volume $V$.
 If the two spheres touch or overlap with each other, they are said
 to be connected and form a part of a cluster.
 Then $N$ and $V$ are taken to infinity with $n = N/V$ being fixed.
 The problem is to find a critical density  $n_c$ at which a
  cluster with infinite size is first formed.
 If one replaces the  spheres here
  by hadrons and $R$ by a typical hadron-radius,
  the close relation between the continuum percolation
 problem and Fig. 1 becomes clear:
 $n_c$ can be interpreted as a critical hadron density at which
 color conductivity starts to be non-vanishing.
 One can also introduce another critical density where the hadrons are
 close-packed. This corresponds to  an end of the percolation where
 the full color conductivity is
realized.

 The continuum percolation problem has been studied using
 numerical simulations and the renormalization group techniques
\cite{PS}.
 The first left column in Table 1 shows the
 result of the numerical simulations of the onset density
 of percolation $n_c$ (multiplied by
 the size of the sphere $v \equiv (4\pi/3) R^3$).
 The close-packed density is, on the other hand, simply defined by $n_c v = 1$.
 In the second column, $n_c$ itself is shown by adopting the
typical hadronic size as $R=0.65 $fm (which is the pion size).
 In the third column, corresponding mean distance between the
 hadrons are shown.  When the percolation starts, the mean distance
 $d_c$ is 3 times  larger than the hadron radius, thus the
 system is still dilute.

\

\vspace{0.8cm}

\noindent
Table 1. Critical values for the continuum percolation in QCD.

\vspace{.5pc}

\begin{tabular}{|c|c|c|} \hline
       & percolation-start & percolation-end  \\
       & (infinite cluster formed) & (close-packed)  \\   \hline \hline
$n_cv$ & $0.35 \pm 0.06$   & 1.0   \\ \hline
$n_c (R=0.65$fm)  & 0.3/fm$^3$   & 0.89/fm$^3$   \\ \hline
$d_c$              & 1.84 fm   &  1.3 fm  \\ \hline
\end{tabular}

\vspace{0.8cm}

Using Fig. 2 and $n_c$ in Table 1, one can estimate
 the critical temperature corresponding to the start (end)
 of the percolation, which results in $T_c \simeq 160\ (180) $ MeV.
 In this region, massive resonances dominate over the pions, thus
 the percolation  transition from the hadronic phase
 to the quark-gluon plasma is likely to occur in the Hagedorn regime.
 One should also note that, in spite of the large difference
 between $n_c$ for start and end of the percolation,
 $T_c$ of the two  are very close. This
 is simply because the resonance contribution to $n$ grows very fast
  as $T$ increases in the Hagedorn regime as can be seen in Fig. 2.
  One should note here that the estimate of $T_c$ above is
 only qualitative: Inclusion of more massive resonances
 will decrease $T_c$ further.

One can also play similar game for the system at
 zero $T$ but at high baryon density. In this case, the
 uncertainty of  the critical baryon density $\rho_c$ is large
but one  gets $\rho_c \simeq $(a few $\sim$ 10)$\rho_0$ with
 $\rho_0 =0.17/$fm$^3$ being the normal nuclear-matter density.

On the basis of these intuitive and simple analyses, one may draw
 a possible  phase diagram of QCD (Fig.3).
\begin{figure}[t]
   \vspace{13pc}
  \caption{Possible QCD phase diagram. The precise shape of the
 critical line and the order of the transition are not known.}
\end{figure}
 The early universe corresponds to the high $T$ but low $\rho$ region
 in the phase diagram.  Roughly $10^{-5}$ seconds after the big bang,
 the universe cools down to about $T \sim$ 100 MeV
 and undergoes a phase transition to the hadronic phase.
 Deep inside the neutron stars and in the possible quark stars,
 low $T$ but high $\rho$ matter could be formed due to the
 strong gravitational pressure.
 One may also reach the high $T$ and/or high $\rho$ region in the laboratories
 by using the relativistic heavy-ion colliders such as RHIC and LHC.

 Despite the considerable efforts so far, high $\rho$ region is not well
 understood because of the complex QCD/hadron dynamics.
 On the other hand, finite $T$ phase transition is
  studied rather well since one can make use of the
 lattice QCD simulations as a useful guide.
  In the next section,
 I will briefly summarize the
 basic knowledge obtained on the lattice for the finite $T$
 phase transition.

\section{Phase transition on the lattice}

The precise determination of the QCD phase transition
 near $T_c$ is
one of the central issues of the recent lattice QCD studies \cite{LAT}.
  As for the
pure gauge system without dynamical fermions ($m_q=\infty$), the center
symmetry
($Z(3)$ in $SU_c(3)$ case) controls the confinement-deconfinement phase
transition.  The effective $Z(3)$ spin model predicts the 1st order transition
and the lattice studies with finite size scaling analyses support
this feature \cite{FINITE}.
  Although it is of 1st order, the transition is much weaker than
that seen before on the smaller lattices.
 Once one introduces dynamical
fermions, $Z(3)$ symmetry is explicitly broken. However, as far as $m_q$ is
large enough, one can still study the phase transition based on this
approximate symmetry.
 On the other hand, in the opposite limit where
 $m_q$ is zero,  chiral symmetry instead of $Z(3)$ symmetry takes place and
 $\langle \bar{q}q \rangle_{_T}$ becomes an order parameter.
 For finite quark masses
 ($m_{u,d} = O(10$MeV) and $m_s=O(200$MeV)),
   chiral symmetry is explicitly broken,
but one can  study the phase transition based
on this approximate chiral symmetry.

 The order of the chiral transition for the realistic quark masses
 are still not known,
  although the chiral transition near $m_q=0\ (q=u,d,s)$ is likely to
 be of 1st order.
 In the Columbia data \cite{COL} with stagerred fermions,
  the chiral transition is not observed for the realistic
 values of $m_{u,d,s}$ , while the recent
 Tsukuba data with Wilson fermions \cite{Kanaya}
 indicate that the transition is of 1st order
 for the realistic
 values of $m_{u,d,s}$.
  Future large scale simulations with finite size scaling analyses
 are called for to settle the issue.

The order of the finite $T$ chiral transition
 is most relevant to the big-bang nucleosynthesis of
  $^9$Be, $^{10}$B and $^{11}$B \cite{Kajino}.
 The spatial inhomogeneity
 due to the bubble formation during the 1st order chiral transition
 can create those relatively heavy elements, while the  standard
 homogeneous model creates
 only 10-100 smaller abundances.

 From the point of view of the laboratory experiments such as the
 relativistic heavy-ion collisions,
 the precise order of the transition is not much relevant because the
 system size is finite. Instead,
  the global behavior of  entropy density and chiral condensate
 as a function of $T$ is rather important for the
 time-evolution of the system as well as the related experimental signals
 of the formation of the quark-gluon plasma \cite{Matsui}.

\subsection{Energy density, pressure and chiral condensate}

 In Fig. 4(a), the lattice data of the
 energy density and pressure of QCD with
 2 light flavors are shown as a function of $\beta = 6/g^2$ \cite{Got}.
 One can clearly see a rapid growth of the energy density ${\cal E}$
 (square in the
 figure), which indicates the liberation of quarks and gluons at
 high $T$.
 ${\cal E}$ and entropy density are known to
  have rapid growth in a  narrow range of temperature
 ($\sim 10$ MeV)  by various numerical simulations on the lattice \cite{NC}.
  On the other hand, the pressure ${\cal P}$ (cross in the figure)
 does not have rapid growth and does not satisfy
 the Stefan-Boltman's law ${\cal E} = 3 {\cal P}$ above $T_c$.
 This may indicate that there  still remain strong and non-perturbative
 interactions between quarks and gluons  above $T_c$.
 Such non-perturbative effect may have close connection with
 the confining nature of the dimensionally reduced effective
 theory of QCD at high $T$ and also with
 the non-perturbative hadronic correlations above $T_c$.
  I will not go into the details of this interesting subject
 due to the limitation of space
 (see \cite{IH} and the references therein for details).

\begin{figure}[t]
   \vspace{14pc}
   \caption{(a) Lattice data for the energy density and pressure
 in 2-flavors ($ma = 0.025$) on $8^3 \times 4$ lattice [14].
 (b) Lattice data for the chiral condensate in (2+1)-flavors
 on $12^3 \times 6$ lattice (taken from [17] and compiled by W. Weise).}
 \end{figure}

In Fig. 4(b), the lattice data for the chiral condensates as a function
 of $\beta$ in (2+1)-flavors are shown \cite{Kogut}.
 One can see that
 (i) a rapid change of  the light quark condensate
  around $T_c$ and (ii) a slow change of the strange condensate
  across $T_c$. Whether they have discontinuity across $T_c$ or not
 is a controversial matter as I have already mentioned.

\subsection{Effective theory interpretation}

As for the behavior of the chiral condensate shown in Fig.4(b), one can
 interpret the results using the effective theories of QCD.
 In Fig. 5,  calculations based on the Nambu-Jona-Lasinio (NJL)
 model \cite{HK1}
 and on the chiral perturbation theory (ChPT) + massive resonances
 \cite{GL}
 are shown.
 In both cases, a rapid change of the light quark condensate
 is seen. In the NJL model, the change is driven by the
 melting of the constituent quark mass. On the other hand,
 in the ChPT + massive resonances,
   the massive states turn out to be more
 important than the pions for chiral restoration,
  which can be seen by comparing the dash-dotted line (pions)
 and the shaded area (massive states + pions).

The light quark  condensate from the massive resonance contributions
 is written as
\begin{eqnarray}
{\langle \bar{q}q \rangle_{_T} \over \langle \bar{q}q \rangle_0}
 = 1 - \int_0^{\infty}dm \
 {\langle \bar{q}q \rangle_m \over \langle -\bar{q}q \rangle_0}
 \ n(m;T)\  \rho(m),
\end{eqnarray}
where $  \langle \bar{q}q \rangle_m $ denotes the expectation
 value of $\bar{q}q$ with respect to a resonance with mass $m$.
 The formula is relevant near $T_c$ and,
 as in the case of eq.(2), the Boltzman suppression by $n(m;T)$
 is compensated by the exponentially growing degeneracy factor
 $\rho(m)$.
At very low temperature, however, the decrease of the condensate
 is dominated by the pions, namely
\begin{eqnarray}
 { \langle \bar{q}q \rangle_{_T} \over  \langle \bar{q}q \rangle_0 }
 = 1 - {T^2 \over 8f_{\pi}^2}  -
  {T^4 \over 384f_{\pi}^4} - \cdot \cdot \cdot \ \  ,
\end{eqnarray}
 for  massless pion \cite{GL}. Fig.5 shows that
 the critical temperature for the chiral transition is likely
 in the Hagedorn regime, which is consistent with the
 argument in section 2 based on the percolation theory.

At present there is no obvious connection between the NJL picture
 (melting of the constituent mass) and the resonance picture (many
 massive resonances) in the description of the chiral transition.
 It is of great interest to find a unified picture of the
 two.
 The quark-hadron duality, which can be seen through the dispersion relation
 in some processes,
 may be relevant to this problem.

\begin{figure}[t]
   \vspace{13pc}
   \caption{ Quark condensates in the NJL model [18] (left)  and the
  chiral perturbation theory + massive resonances [5] (right).}
\end{figure}

\section{Dynamical critical phenomena}

Since $\langle \bar{q}q \rangle_{_T}$ is a scale dependent quantity, it
 is not a direct experimental
 observable. Therefore, one has to look for other physical quantities
   to see the signal of the chiral
 phase transition in the real experiments.
   The spectral change of hadrons  at finite $T$
 such as the mass shift  is one of the
 possible candidates.
 In fact, light-hadron masses are essentially determined
 by the quark condensates as QCD sum rules tell us \cite{SVZ}, which suggests
 that the mass shift of hadrons in medium  could be a good measure of the
 partial restoration of chiral symmetry at finite $T$ and $\rho$
 \cite{Pis,HK2,BR}.

  There actually exist similar situations in condensed matter physics:
 the existence of the soft  phonon modes
 is an indication that the ground state
  undergoes  structural phase transition and
  one can study the precise nature of the
 phase transition
 by the soft mode spectroscopy (see subsection 4.3 for the examples).
 In QCD, scalar mesons (fluctuation of the order parameter)
 and the vector mesons (such as $\rho$, $\omega$ and $\phi$)
 are the candidates for the  ``soft modes''.
 In particular, if the vector mesons are the soft modes, one can
 directly detect the spectral changes through the leptonic decays
  ($\rho, \omega, \phi \rightarrow e^+e^-, \mu^+ \mu^- $).\footnote{
The spectrum of heavy vector mesons such as $J/\psi$ is  also interesting
 in relation to the physics of confinement and the signature of the
 formation of the quark-gluon plasma \cite{miyamura}.
 However, we will not discuss it in this article.}

Throughout the following sections, I will define the mass of the hadrons
 as a peak position of a resonance in the spectral function.
 Before discussing  the details of the
  dynamical critical phenomena,
 let us first summarize what is know about the static critical phenomena
 in the next subsection.

\subsection{Static fluctuation of the order parameter}

The left two figures in Fig. 6 show schematic illustration of the behavior
 of the chiral condensate $\langle \bar{q} q \rangle$ and
 its static fluctuation
\begin{eqnarray}
 \chi \sim \langle (\bar{q} q)^2 \rangle
 - \langle \bar{q} q \rangle^2,
\end{eqnarray}
 for QCD with  massless two-flavors.
 (In this case
 the phase transition is expected to be of 2nd order.
 Even for the realistic (2+1)-flavors, the
 similar argument holds  as far as the chiral condensate has
 significant variation near $T_c$.)
 Precisely at the temperature where $\langle \bar{q} q \rangle$
 vanishes, $\chi$ is expected to diverge.
 The right two figures of Fig. 6 are the corresponding lattice
 data \cite{Got88,Kar94}.
 ``disc'' in the lower right figure corresponds to $\chi$ which
 clearly
 shows a large enhancement near $T_c$
  for light quark ($ma=0.02$). This confirms
 our theoretical expectation.
  On larger lattices, one will be able to extract
   the
 static critical exponents accurately \cite{Kanaya2}.

Although these static quantities  are by themselves interesting,
 our main concern here is the quantity which can be measured in
 the laboratories.   I will thus concentrate on the
 time-dependent (dynamical) phenomena in the following sections.

\begin{figure}[t]
   \vspace{22pc}
   \caption{ Left two figures: Theoretical expectation for the chiral
 condensate and the static susceptibility in the 2nd order chiral transition.
 Right two figures: Corresponding lattice data (upper: $N_f=2$, $8^3 \times
 4 $ lattice with $ma=0.025$  [24],
 lower: $N_f=2$, $8^3 \times 4$ lattice with $ma=0.02,0.0375,0.075$ [25].)}
\end{figure}

\subsection{Dynamical fluctuations and para-pion}

A possible dynamical mode which has similar critical behavior
 with the condensate is the time-dependent fluctuation
 of the order parameters dictated by the retarded correlation function
\begin{eqnarray}
\Pi(\omega,{\bf q})
 = i \int d^4x \ e^{iqx}\theta(x^0) \langle [\bar{q}\Gamma q(x),
 \bar{q} \Gamma q(0)] \rangle_{_T} ,
\end{eqnarray}
where
 $\Gamma$ denotes $1$ or $i \gamma_5 \tau^a$.
 Expected behavior of the ``mass'' (peak position of the
 spectral function  ${\rm Im} \Pi(\omega,{\bf q})$) is shown
 in the left hand side of Fig. 7.
 The scalar mode ($\Gamma =1$) corresponds to $\sigma$ and the
 pseudo-scalar mode ($\Gamma =i \gamma_5 \tau^a$)
 corresponds to $\pi$.  The right figure is a calculation
 using the NJL model with $m_{u,d}({\rm 1GeV})=5.5$MeV \cite{HK3}.

\begin{figure}[t]
   \vspace{15pc}
   \caption{Left: Theoretical expectation for the masses of the
 dynamical modes $\sigma$ and $\pi$. Right: Calculation using the
 NJL model with $N_f=2$ and  $m_{u,d}(1{\rm GeV}) = 5.5 MeV$ [27].}
\end{figure}

 From Fig.7, one confirms the naive expectation that
 the chiral multiplet will degenerate at and above $T_c$ and also
 one can learn more:

\noindent
\ \ \ \ (i) There is a sizable softening (decreasing mass) of $\sigma$
 below $T_c$. This is an indication that the
 ground state is soft for the deformation to the
 direction of the order parameter.
  One can also calculate the decay width $\sigma \rightarrow 2 \pi$
 at finite $T$.  Just because of the tendency $m_{\sigma} \rightarrow
 m_{\pi}$, the decay width is suppressed near $T_c$ \cite{HK3}.
 However, whether $\sigma$ becomes really a sharp resonance
 or not depends on the magnitude of the collisional width
 which has not been calculated for $\sigma$  so far.

\noindent
\ \ \ \ (ii) Above $T_c$, $\sigma$ and $\pi$ are degenerate and
 have low mass.  Furthermore, the decay channel
 $(\sigma,\pi) \rightarrow \bar{q}q$ near $T_c$ is suppressed
 by the small phase space, therefore the width of the modes are small.
 In Fig. 8, shown is the spectral function of this degenerate mode
 above $T_c$ calculated by using the NJL model \cite{HK2}.
 We will call this mode as {\bf para-pion} since it
 can be regarded as a low-mass and long-lived elementary excitation
 in the para-phase of chiral symmetry.

\begin{figure}[t]
   \vspace{15pc}
   \caption{Spectral function for
 para-pion above $T_c$ in the NJL model [21].}
\end{figure}

\subsection{Soft modes -- examples in solid state physics --}

 The softening associated with the ferro-para phase transition is
 well known in sold state physics and  the dynamical excitations
  softened near the critical point are
 generally called {\bf soft modes} \cite{Soft}.
In Fig. 9, one of such examples
 is shown. The figure shows a soft mode in the
 ferro-electric crystal where the self-polarization
 $\vec{P}$ is an order parameter \cite{FE}.
  Similar soft mode can be also seen
 in the ferro-elastic crystal where the self-distortion $\vec{X}$
 is an order parameter \cite{FEL}.

\begin{figure}[t]
   \vspace{15pc}
 \caption{Soft mode in the ferro-electric crystal [29].}
\end{figure}

\section{Vector mesons at finite T}

 Although the softening of the scalar meson $\sigma$ below $T_c$
 and the existence of para-pion above $T_c$ are the interesting
 dynamical critical phenomena, it is rather difficult to measure
 them  in the relativistic heavy ion collisions. The reason is that
 the hadronic decay of $\sigma$ is  masked by thousands of
  pions produced by the collisions.  There might be a chance,
 however, to see the decay $\sigma \rightarrow 2 \gamma$
 which occurs when the system is at finite density \cite{Wel}.

 On the other hand, the light vector mesons such as $\rho$, $\omega$ and
 $\phi$ are more interesting from the experimental point of view.
 They decay into lepton pairs ($e^+e^-$ and $\mu^+ \mu^-$) which
 can penetrate the hot/dense medium without strong interactions.
 Thus, if vector mesons are the soft modes (which is not obvious
 from the outset), the lepton pairs are the good probe to see the
 dynamical critical phenomena.
 To study this possibility, let's start with the retarded hadronic
 correlation at finite $T$ in the vector channel:
\begin{eqnarray}
\label{start}
 \Pi^V_{\mu \nu} (\omega,{\bf q})
 = i \int d^4x \ e^{iqx} \theta(x^0) \langle [J^V_{\mu}(x),J^V_{\nu}(0)]
 \rangle_{_T} ,
\end{eqnarray}
where $J^{\rho, \omega}_{\mu} = \bar{u}\gamma_{\mu}u \mp \bar{d}\gamma_{\mu}d$
 and $J^{\phi}_{\mu}=\bar{s}\gamma_{\mu}s$.
  $\Pi^V (\omega) \equiv \Pi^V_{\mu \mu} (\omega,{\bf q}=0)/(-3 \omega^2)$
  satisfies the following dispersion relation
\begin{eqnarray}
 \Pi^V (\omega)
 = {1 \over \pi} \int_0^{\infty}
{{\rm Im} \Pi^V(\omega')  \over \omega'^2-(\omega+i\epsilon )^2 }
 d\omega'^2 + ({\rm subtraction}).
\end{eqnarray}
What we are interested in is the spectral change of
 ${\rm Im} \Pi^V$ at finite $T$.

\subsection{Low T theorem in the Yukawa regime}

  At extremely low temperature,
 spectral changes of hadrons  are   controlled solely
 by chiral symmetry.  In fact, the forward scattering
 of thermal pions by a hadron
is only the relevant process to change the
properties of the hadron at low $T$.
  Leutwyler and Smilga  have
shown that the masses of light hadrons (nucleon, $\rho$-meson etc)
 do not  change at  $O(T^2)$, though
 the pole residues can be modified  at this  order \cite{LS}.

 For the $\rho$ meson, the spectral change at $O(T^2)$
 can be easily calculated by using the soft pion theorem \cite{DEI};
\begin{eqnarray}
\label{lowT}
{\rm Im} \Pi^V(\omega) \simeq (1-{T^2 \over 6f_{\pi}^2})
 {\rm Im} \Pi^V_{T=0}(\omega)
+{T^2 \over 6f_{\pi}^2}
 {\rm Im} \Pi^A_{T=0}(\omega),
\end{eqnarray}
 where  $ \Pi^{V,A}_{T=0}(\omega)$ denotes the correlation function
 at zero $T$ in the vector ($V=\rho$-meson) and axial-vector ($A=a_1$-meson)
 channel.
 It is obvious from the formula that the thermal pions
 induce the mixing between V and A channels and also modify the
 pole residues, but do not change the mass.
 See   Fig.10(a) for the physical process to induce the V-A mixing.

\begin{figure}[t]
   \vspace{10pc}
  \caption{(a) Mixing between $\rho$ and $a_1$ mesons through thermal pion.
 (b) Mixing  between $\rho$ and other hadrons through thermal resonances.}
\end{figure}

\subsection{Approaching $T_c$ -- Hagedorn regime --}

As we have seen in Section 2, the real interesting region is in the
 Hegedorn regime where resonances dominate over pions.
 In this regime, the vector mesons will have
 interactions with various thermal resonances and the final states
 are not limited to the axial vector meson. See Fig. 10(b).
 In this complex situation, there are two possible ways to study
 the spectral change.

\noindent
\ \ \ \ (i) Lattice QCD: By measuring the imaginary time correlation
 on the lattice, one can in principle reconstruct
 ${\rm Im} \Pi^V(\omega) $ through the dispersion relation \cite{Hashi}
\begin{eqnarray}
 \Pi^V (i\omega_n)
 = {1 \over \pi} \int_0^{\infty}
{{\rm Im} \Pi^V(\omega')  \over \omega'^2+\omega_n^2 }
 d\omega'^2 + ({\rm subtraction}),
\end{eqnarray}
where $\omega_n = 2 n \pi T$ is the Matsubara frequency.

\noindent
\ \ \ \ (ii) QCD sum rules: By calculating the real time correlation in the
 deep Euclidian region using the operator product expansion,
 one can in principle reconstruct
 ${\rm Im} \Pi^V(\omega) $ through the dispersion relation \cite{BS,HKL}
\begin{eqnarray}
\label{dispe}
 {\rm Re}\Pi^V (\omega^2 \rightarrow - \infty)
 = {1 \over \pi} \int_0^{\infty}
{{\rm Im} \Pi^V(\omega')  \over \omega'^2-\omega^2 }
 d\omega'^2 + ({\rm subtraction}).
\end{eqnarray}

The lattice approach is still far from being realistic because the
 temporal lattice size is currently too small.  We will
 thus pursue the QCD sum rules in the following
 to get some constraints on the spectral function.

One should note here that the hadronic screening mass defined
 by the ``spatial'' hadronic correlation does not have direct
 connection with the ``real mass" which we are working on  here.
 The calculation of the screening mass
 on the lattice is, however, much easier than the real mass and
 it has been and is being studied extensively  \cite{DeTarKogut}.

\subsection{QCD sum rules at finite $T$}

QCD sum rules in medium \cite{HKL}
 start with the following operator product expansion (OPE)
 for $\Pi^V(Q^2)$ with $Q^2 \equiv - \omega^2 $;
\begin{eqnarray}
\label{OPE}
{\rm Re} \Pi^V (Q^2) = - C_0 \ln Q^2 + \sum_{n=1}^{\infty}
 {C_n \over Q^{2n}} \langle {\cal O}_n \rangle_{_T} ,
\end{eqnarray}
where $C_n$ are the c-number Wilson coefficients which are
 $T$ independent. All the medium effects are in the
 thermal average of the local operators ${\cal O}_n$.
 Since $\langle {\cal O}_n \rangle_{_T} \sim T^{2l}\cdot \Lambda_{QCD}^{2m}$
 with $l+m=n$ due to the dimensional reason, (\ref{OPE})
 is a valid asymptotic expansion as far as
 $Q^2 \gg T^2, \Lambda_{QCD}^2$.

The first 4-terms of $C_n \langle {\cal O}_n \rangle_{_T} $ has been
 calculated as \cite{HKL}
\begin{eqnarray}
C_0 & = & - {1 \over 8 \pi} (1+ {\alpha_s \over \pi}) , \ \ \
 C_1  =  0 ,  \\
\label{dim4}
C_2 \langle {\cal O}_2 \rangle_{_T} & = & {1 \over 24}
\langle {\alpha_s \over \pi} G^2 \rangle_{_T} +
 {4 \over 3} \langle {\cal S} \bar{q} i \gamma_0 D_0 q \rangle_{_T} , \\
C_3 \langle {\cal O}_3 \rangle_{_T} & = & -
 \langle {\rm scalar\ 4-quark}) \rangle_{_T}
 +
 {16 \over 3} \langle {\cal S} \bar{q} i \gamma_0 D_0 D_0 D_0 q \rangle_{_T} .
\end{eqnarray}
Here we have neglected the terms proportional to the
 light quark masses and the quark-gluon mixed operators.
 Also, ${\cal S}$ makes the operators symmetric and traceless.
At low $T$ (Yukawa regime), one may use the soft pion theorems
 and the parton distribution of the pion to estimate the r.h.s.
 of the above equations.
 When $T$ is close to $T_c$, one has to look for totally different
 way of estimation: a simplest approach is to assume the resonance gas
 to evaluate the r.h.s., while the direct lattice simulations
  will be the most reliable way in the future.

 An important feature of the OPE in the above is that
  local operators with Lorentz indices arise.  This happens because
 we are taking the rest frame of the heat bath which breaks
  covariance.  To see the effect of the operators with
 Lorentz indices, let's look at the $T$ dependence of
 the r.h.s. of  (\ref{dim4}) using
 the resonance gas approximation of $\pi$, $K$ and $\eta$.
 As is seen from Fig.11,  these resonances
 increase the second term of the  r.h.s. of (\ref{dim4})
 considerably, which acts to reduce the mass of the vector mesons as
 we will see later.

\begin{figure}[t]
   \vspace{15pc}
   \caption{$C_2 \langle {\cal O}_2 \rangle_{_T}$ with ($a+b+c)$
 and without ($a+b$) the
 contribution $\langle \bar{q} i \gamma_0D_0q \rangle_{_T} $ in the resonance
 approximation [36].
The solid (dashed) line includes the thermal
 contribution of $\pi$ ($\pi$, $K$, $\eta$).}
\end{figure}

Using  OPE given in (\ref{OPE}) and the dispersion relation (\ref{dispe}),
 one can  construct the following energy weighted  sum rules
 or the finite energy sum rules (FESR) \cite{HKL};
\begin{eqnarray}
I_1 & = & \int_0^{\infty}
 [{\rm Im} \Pi^V(\omega) - {\rm Im} \Pi^V_{pert.}(\omega)] d\omega^2 = 0,
  \\
 I_2 & = & \int_0^{\infty}
 [{\rm Im} \Pi^V(\omega) - {\rm Im} \Pi^V_{pert.}(\omega)]\omega^2
 d\omega^2 = - C_2 \langle {\cal O}_2 \rangle_{_T}, \\
I_3 & = & \int_0^{\infty}
 [{\rm Im} \Pi^V(\omega) - {\rm Im} \Pi^V_{pert.}(\omega)] \omega^4
 d\omega^2 = C_3 \langle {\cal O}_3 \rangle_{_T} .
\end{eqnarray}
Here ${\rm Im} \Pi^V_{pert.}(\omega)$, which is $T$ independent,  denotes the
 imaginary part corresponding to the perturbative part
 of $\Pi^V$.  Similar sum rules hold  for the axial vector
 channel (in the chiral limit)
 except that one has a different operator for ${\cal O}_3$.
 One can also generalize the above  sum rules to finite
 ${\bf q}$ \cite{KS}.

We have only three constraints $I_{1,2,3}$ from the QCD sum rules
 as shown above.
 This is because we could calculate only first three non-perturbative
 terms $C_1 {\cal O}_1$, $C_2 {\cal O}_2$ and $C_3 {\cal O}_3$ in OPE.
 By the three  constraints, we can determine three
 resonance parameters but not more than three. In this situation,
 the  most economical way to parametrize
 the spectral function ${\rm Im}\Pi^V(\omega)$ is to introduce
 the following three parameters; the position of the resonance peak $m(T)$,
 the continuum threshold $S_0(T)$ and
 the area integral of the resonance
 $F(T) = \int_{0+}^{S_0} {\rm Im} \Pi^V(\omega) d \omega^2$.\footnote{
 ${\rm Im} \Pi^V(\omega)$ has a Landau-damping contribution
 at $\omega =0$ which we have calculated explicitly
 in the resonance gas approximation and is not included in $F(T)$
 \cite{BS,HKL}.}

 With three sum rules and the parametrization of
 ${\rm Im} \Pi^V(\omega)$ in the above, it is impossible to extract the width
 of the resonance.
  A honest way to determine the width is to go to dim. 8 operator
  $C_4 {\cal O}_4$ and derive one more sum rule, which is practically
 a formidable task.

\subsection{Spectral change in Yukawa regime and Hagedorn regime}

\vspace{0.2cm}

\noindent
{\underline{Yukawa regime}}:

By examining the FESR or the Borel sum rules at low $T$, one can show
 the followings.

\noindent
\ \ \ \ (i) The low $T$ theorem (\ref{lowT}) is satisfied
  \cite{HKL,EI94};\footnote{
 The difference of the OPE in vector channel and the
 axial-vector channel appears only in the dim.6 operator \cite{HKL}.
 At low $T$, this difference is completely absorbed by the
 $V-A$ mixing and does not cause mass shift, which can be
 seen explicitly \cite{EI94} or numerically \cite{HKL}
 in QCD sum rules.  The similar situation also arises
 for the nucleon mass at $O(T^2)$ \cite{koikeN}.}
\begin{eqnarray}
\delta m_{V,A} = 0 \ \ \ {\rm at} \ \ \ O(T^2).
\end{eqnarray}

\noindent
\ \ \ \ (ii) Negative mass shifts occur at $O(T^4)$ \cite{EI94};
\begin{eqnarray}
\delta m_{V} \sim  \delta m_{A} = - c T^4 \ \ \ {\rm at} \ \ \ O(T^4),
\end{eqnarray}
with a small positive coefficient $c$.

\vspace{0.2cm}

\noindent
{\underline{Hagedorn regime}}:

When $T$ is close to $T_c$,
 the sum rules are not powerful enough
 to predict the precise spectral change. In fact, there are
  two physical possibilities.

\noindent
\ \ \ \ CASE-I: The widths of the hadrons increase rapidly and
 no distinction between the resonance and the continuum
  can be made near $T_c$.

\noindent
\ \ \ \   CASE-II: There still have clear distinction between the
 continuum and the lowest resonance near $T_c$.

Since CASE-II is a more interesting possibility experimentally and
 also CASE-II is actually realized in many solid state examples,
 let's focus our attention on it for the moment and
 try to see what kind of constraint one can make from the QCD sum rules.
 In Fig.12, the mass and the continuum threshold for the $\rho, \omega$
 mesons
 at finite $T$ are shown. The calculation was made using the Borel version
 of the QCD sum rules (FESR gives essentially the same results).
 The curves (a) and (b)
correspond to
 the assumptions (a) and (b)  in
 Table 2 respectively.

 The decreasing mass and threshold are  clearly induced by the
 dimension 4 operators, and if dimension 6 operator decreases
 at finite $T$ as (b), it enhances the decrease further.
 Note here that  $\langle {\cal O}_3 \rangle_{_T}$
 in (b) is simply an ansatz:  near $T=0$, more rigorous
 $T$ dependence is know  for this quantity \cite{HKL}.

\vspace{0.5cm}

\noindent
Table 2. Assumed dimension 4 and 6 condensates in Fig. 12.

\vspace{.5pc}

\begin{tabular}{|c|c|c|} \hline
       & $\langle {\cal O}_2 \rangle_{_T}$ &$\langle {\cal O}_3 \rangle_{_T}$
  \\
\hline \hline
(a)  & $\pi, K, \eta$ resonance gas &  no $T$-dependence  \\ \hline
(b)  & $\pi, K, \eta$ resonance gas & mean-field ansatz
$ \sim (1-T^2/T_c^2)^{1/2}$  \\ \hline
\end{tabular}

\vspace{0.5cm}

\begin{figure}[t]

\vspace{15pc}

   \caption{Temperature dependence of the $\rho, \omega$ mass and the
continuum threshold $S_0$ in the Borel sum rule in CASE-II.
 The upper (lower) curves for $m_{\rho,\omega}$ and $S_0$ correspond
 to the case (a) ((b)) in Table 2. }
\end{figure}

\subsection{$\phi$ meson near $T_c$}

As we have seen in Fig.4(b), the strangeness condensate decreases
more slowly than the u-d condensate at finite $T$.
 Nevertheless, one could probe the decrease through the
 spectral change of the $\phi$ meson:
 Since $\phi$ width is rather small in the vacuum (4.5 MeV)
 and will not increase more than 25 MeV even at $T \simeq 180 MeV$ \cite{Hag},
 the experimental uncertainty for detecting the spectral change
 of $\phi$  will be  less than
 that for $\rho$.
 Asakawa and Ko have generalized the method of  \cite{HKL}
 and calculated the $\phi$ mass at finite $T$
 in CASE-II in the  resonance gas approximation of
  $\pi, \rho,\omega,K,K^*,\eta,N,\Delta,\Lambda, \Sigma$ \cite{AK}.
 As is shown in Fig. 13, $\phi$ mass decreases, which is mainly
 induced by the decrease of the dimension 4 condensate
 $m_s \langle \bar{s}s \rangle_{_T}$.
 The observable consequence of this shift is the double $\phi$ peak
 in the $e^+e^-$ spectrum at RHIC and LHC
  which I will come back in section 7.

\begin{figure}[t]
   \vspace{13pc}
  \caption{$\phi$-meson mass as a function of $T$
 in the QCD sum rules with resonance gas approximation [42].}
\end{figure}

\subsection{Collision widths of vector mesons}

 Since QCD sum rules do not tell us the change of the
  hadronic widths,  it will be desirable to see how large the
 widths could be at finite $T$ using other method.
 Haglin has recently evaluated the collisional widths of $\rho,\omega$
 and $\phi$ through the process
\begin{eqnarray}
 r + V \rightarrow {\rm (two-body\  final\  states)},
\end{eqnarray}
 where $V$ denotes the vector mesons and $r$ is the thermally excited
 resonances ($\rho,\omega,\pi,K,K^*,\phi$ are taken into account
 in the calculation) \cite{Hag}.

\begin{figure}[t]
   \vspace{14pc}
 \caption{ Collision width of  $\phi$ in the
 resonance gas of $\rho, \omega, \pi, K, K^*, \phi$ [41].
}
\end{figure}

 Fig. 14 shows that  $\rho$ and $\omega$ acquire 50-100 MeV width
 near $T =180$ MeV, while $\phi$ acquires at most 25 MeV width.
 These widths are comparable or even larger than
 the widths in the vacuum ($\Gamma_{\rho}=150$ MeV, $\Gamma_{\omega}
 = 8.5 $ MeV and $\Gamma_{\phi} = 4.5$ MeV).
 Nevertheless, the total width is still small for
 $\omega$ and $\phi$ and they can be considered as  distinct
 resonances. $\rho$ is marginal in this respect. However,
 the negative mass shift of $\rho$ (which is not considered
 and is not attainable in Haglin's kinetic theory approach)
 has an effect to reduce the width
 because of the phase space suppression of $\rho \rightarrow 2 \pi$.
 Thus there is still a possibility to have   rather distinct
 $\rho$ if its mass has downward shift.

\subsection{$T$-dependent Wilson coefficients -- Are they meaningful ? --}

In the early stage of the application of the QCD sum rules in the medium
 \cite{BS,DN}
 there was a confusion on the OPE at finite temperature:
 the thermal quark propagator was used to calculate the
 Wilson coefficients of OPE series, which gives $T$-dependent
 Wilson coefficients.
 Although it was later clarified that such procedure does not make sense
   \cite{DEI,HKL,EI94} and  correct sum rules have been developed \cite{HKL},
 the confusion is still floating around in the literatures (see e.g.
 \cite{Dom}).
 So I will make a few remarks on this point here.

\noindent
\ \ \ \ (i) First of all, one should remember that
 OPE of $J(x)J(0)$  is
 based on the factorization of the soft scale and hard scale \cite{muta}.
  The soft dynamics  is in the matrix elements of local operators
 while the hard dynamics is in the Wilson coefficients which
 are by definition  independent of the states
  sandwiching the current product $J(x)J(0)$ (target independence
 of the Wilson coefficients).
 This factorization property is a basis of the whole  success
  of the deep inelastic scattering physics.
   In our case, the hard scale to be taken into account in the
 Wilson coefficients   is $Q$ and the  soft scales to be taken
 into account in the matrix elements are $T (< $250MeV) and
  $\Lambda_{QCD} (\sim $ 200MeV).
  As fas as one keeps $Q \gg  T, \Lambda_{QCD}$, OPE is unique and
 there is no room to have $T$-dependent Wilson coefficients.

\noindent
\ \ \ \
(ii) It is physically
 erroneous to use free thermal quark-propagator in  calculating
 the correlation function below $T_c$.
   Since quarks are strongly interacting and confined below $T_c$,
  there is no reason to believe that the quarks have free
 thermal distribution even close to  $T_c$.

\noindent
\ \ \ \ (iii) The statement (i) in the above has nothing to do with the
 complete set $\mid l \rangle$ one adopts to evaluate the thermal
 average of $\sum_l \langle l \mid J(x)J(0) \mid l
 \rangle \exp (-E_l/T)$ in (\ref{start}).
  In other words, eq.(\ref{OPE}) is  an  exact expression
 when  $Q \gg  T, \Lambda_{QCD}$.  How to calculate
 $\langle {\cal O}_n \rangle_{_T}$ is a different  problem: the direct
 lattice QCD simulation will become the best way to estimate the
 magnitude in the future.  At present, what one can do at best is to
 assume resonance gas of hadrons near $T_c$
 to estimate  $\langle {\cal O}_n \rangle_{_T}$.

 In conclusion, the answer to the headline of this subsection is NO,
 and  analyses based on the $T$-dependent Wilson
 coefficients are not justified.

\subsection{Remarks on the effective theory approaches}

 There exist many attempts so far to calculate the
 spectral change of the vector mesons using different versions of the
 gauged non-linear (and linear) $\sigma$-models \cite{model}.
 In most of these models, the basic ingredients in the Lagrangian
  are limited to the vector mesons and pions (and $\sigma$ meson).
 Since the pion is the lightest particle, it is dominantly excited
 at low temperature and continued to be dominant even at high $T$
 since there are no other possible resonances in the lagrangian.

 As we have already shown in detail
in section 2 (see e.g. Fig.2), the description of the hot matter
 using pion alone is inadequate for chiral restoration and
 massive resonance contribution is inevitable to get
 realistic value of $T_c$.  In this respect, the above model
 calculations, which shows increasing (decreasing) $\rho$ ($a_1$) mass
 and constant $\omega$ mass are valid only at low temperature
 (Yukawa regime) and totally different method is required to
 predict the mass shift near $T_c$ (Hagedorn regime).

 A possible way to incorporate the higher resonance contribution
 effectively is to use the chiral quark model \cite{GM} where
 vector mesons  are interacting with constituent quarks
 with mass $M \simeq 350$ MeV.
 In such models, the increasing number of resonances and
 the simultaneous decrease of the quark condensate at finite $T$
 are simulated by the decreasing $M(T)$.
 Then one can make a physical argument
 that
  the vector meson mass decreases as far as $M(0) > M(T)$ \cite{KH95}.
 Another effective theory predicting the decreasing
 vector meson mass is the Brown and Rho's lagrangian \cite{BRG}
 based on  the  Georgi's vector symmetry \cite{GEOR}.
 The relation of this approach to the previous one is not known however.

\section{Partial restoration of chiral symmetry at finite density}

In the preceding sections, we have concentrated on the
 hadron properties at finite $T$. What we found is that
 the hadronic mass shift as well as the change of the chiral condensate
 occur only when $T$ is close to  $T_c$.
 At finite baryon density, the situation is quite different and one may
 expect partial restoration of chiral symmetry even
 in the heavy nuclei. The basis of this assertion is that
 the quark condensate calculated in the  Fermi gas approximation
 decrease considerably in  nuclear matter \cite{DL,HHP}.
\begin{eqnarray}
\label{condrho}
{\langle \bar{u}u \rangle_{\rho} \over \langle \bar{u}u \rangle_0}
 = 1- {4\Sigma_{\pi N}\over f_{\pi}^2m_{\pi}^2}
 \int^{p_{_F}} {d^3p \over (2\pi)^3} {m_N \over E_p} ,
\end{eqnarray}
where $E_p \equiv \sqrt{p^2 + m_N^2}$, $\Sigma_{\pi N} = (45 \pm 10) $MeV,
 and the integration for $p$ should be taken
  from 0 to the
 fermi momentum  $p_{_F}$($p_{_F}=270$ MeV at normal nuclear matter
 density $\rho_0 = 0.17/{\rm fm}^3$).
 At $\rho=\rho_0$,
 the above formula gives (34$\pm$8)\% reduction of the
 chiral condensate from the vacuum value. See Fig.15.
  The corrections to this simple fermi-gas approximation
 arise in higher orders in $p_F$. Several estimates show that
 such corrections are small at $\rho_0$ \cite{Ko94}.
 If this is the case, the heavy nuclei is a good testing ground of
 the partial chiral restoration: the vector mesons in nuclei
 will be the best probe to see the effect.

\begin{figure}[t]
   \vspace{13pc}
\caption{Condensates vs $\rho/\rho_0$ in nuclear medium:
 $S = \langle \bar{s}s \rangle_{\rho}/\langle \bar{s}s \rangle_{0}$,
 $U = \langle \bar{u}u \rangle_{\rho}/\langle \bar{u}u \rangle_{0}$
 and
 $G = \langle G^2 \rangle_{\rho}/\langle G^2 \rangle_{0}$.
 As for the OZI violation in the nucleon
 $y = 2 \langle \bar{s}s \rangle_{N}/ \langle
  \bar{u}u + \bar{d}d \rangle_{N} = 0.12 $ is used.
 This figure is taken from the third reference in [1].}
\end{figure}

\subsection{Vector mesons in nuclear matter}

Let's first
 consider $\rho$ and $\omega$ mesons propagating inside the nuclear matter.
 Adopting the same fermi-gas approximation with (\ref{condrho}) and taking
 the vector meson at rest (${\bf q}=0$)  for simplicity,
 one can generally write the mass-squared shift (or the self-energy)
 $\delta m^2_{_V} \equiv m^{*2}_{_V} - m^2_{_V}$ as
\begin{eqnarray}
\label{massshiftf}
\delta m^2_{_V}  = 4 \int^{p_f}
  {d^3 p \over (2 \pi)^3  }  {m_{_N} \over E_p}  f_{VN}({\bf p}),
\end{eqnarray}
where $f_{VN}({\bf p})$ denotes the vector-meson (V) -- nucleon (N)
 forward scattering amplitude
 in the relativistic normalization  (see Fig.16).
 Here, we took spin-isospin average for the nucleon states
 in $f_{VN}$; that is why we have a degeneracy factor 4 in the r.h.s.
 of (\ref{massshiftf}).
The integration for $p$ should be  taken up to the fermi momentum $p_{_F}$.

\begin{figure}[t]
   \vspace{11pc}
\caption{Forward scattering of vector meson with the nucleon in nuclear
 matter.}
\end{figure}

If one can calculate $f_{VN}({\bf p})$ reasonably well in the range
 $0 < p < p_{_F}=270$ MeV
 (or $1709\ {\rm MeV} < \sqrt{s} < 1726\ {\rm MeV} $ in terms of the
 $V-N$ invariant mass), one can predict the mass shift.
 Unfortunately, this is a formidable task.
 First of all, $f_{VN}({\bf p})$ is not a constant at all
  in the above range since there are at least two
 s-channel resonances $N(1710), N(1720)$ in the  interval
 and two nearby resonances $N(1700)$ and $\Delta(1700)$.
  They all couple to the $\rho-N$ system \cite{PDG}
 and give a rapid variation
 of $f_{VN}({\bf p})$ as a function of $p$.
 Secondly,
 there  are t-channel meson exchanges between $\rho$ and $N$, which are
  difficult to estimate since we do not know what are
 the relevant mesons and what are their
 couplings with $\rho$ and $N$.
 Even if one could manage these s-channel and t-channel contributions
 assuming some effective lagrangian, the connection of the resulting
 forward amplitude and the chiral condensate is still missing.
 Therefore one should look for totally different approach to estimate
 $\delta m_V$, one of which is the QCD sum rules in medium
 discussed in the
 following subsections.

\subsection{Constraints from QCD sum rules}

Starting from the retarded correlation of the vector currents
 in nuclear medium, one can write down the
 following FESR constraints \cite{HL}.
\begin{eqnarray}
\label{finitedsum}
\int_0^{\infty} [{\rm Im}\Pi^V(\omega) - {\rm Im}^V\Pi_{pert.}(\omega)]
 \omega^{2(n-1)} d\omega^2 = (-)^{n-1}C_n \langle {\cal O}_n \rangle_{\rho},
\end{eqnarray}
 with $n=1,2,3$.
 In the  r.h.s., the matrix elements of the local operators
 are calculated in the fermi gas approximation.
 If one introduces three parameters as before, namely
 the peak position of the resonance,
 the continuum threshold and the integrated strength of the resonance,
  one can extract the
 density dependence of these parameters from (\ref{finitedsum}).
 In Fig. 17, results of  such analysis using the Borel sum rule
 are shown  (FESR gives essentially the same result) \cite{HL}.
  By making the linear fit, one could deduce
 a formula for the mass shift as
\begin{eqnarray}
{m_{_V}^* \over m_{_V}} \simeq 1 - c_{_V} {\rho \over \rho_0} ,
\end{eqnarray}
where $c_{\rho, \omega} = 0.18 \pm 0.05$ and
 $c_{\phi} = (0.15 \pm 0.05)  y$ with $y = 0.1 - 0.2$ being the
 OZI breaking parameter in the nucleon (see the figure caption of Fig.15
 for the definition).
 These numbers are obtained by neglecting the contribution of the
  quark-gluon mixed operator with twist 4; inclusion of them
  moves the central value of $c_{\rho,\omega}$ to $0.15$ \cite{HLS}.

 The scaling argument of Brown and Rho \cite{BR} and the
 Walecka model of nuclear matter \cite{HS} also
 predict the similar decrease of  the $\rho$ and $\omega$ masses.

\begin{figure}[t]
   \vspace{16pc}
 \caption{(a) Masses of $\rho$, $\omega$ and $\phi$ mesons
 in nuclear matter predicted in the QCD sum rules  (left) [55]
 together
 with the prediction of the Walecka model (right) [57]. $M^*/M$
 in the right figure shows the effective mass of the nucleon.}
\end{figure}

\subsection{Use and misuse of the QCD sum rules in nuclear medium}

 Let me comment  more
 on the formula (\ref{massshiftf}), since it
  has created confusion
 in the literatures (see e.g. \cite{Koike})
  on the vector mesons in nuclear matter.

  As I have mentioned, $f_{VN}({\bf p})$ must be  a rapidly varying
 function of $p$. Thus it is impossible to approximate
 it by $f_{VN}(0)$:
\begin{eqnarray}
\label{nono}
 f_{VN}({\bf p}) \neq f_{VN}(0)  \  \ \ \ \ {\rm except \ for} \ \ p\sim 0 .
\end{eqnarray}
 Note that the $V-N$ scattering length $a_{_{VN}}$ is proportional to
 $f_{VN}(0)$.

In terms of the mass shift,
 the (invalid) approximation $f_{VN}({\bf p}) = f_{VN}(0)$
 for $0 < p < p_F$
 implies that
\begin{eqnarray}
\label{masss}
 \delta m^2_{_V} \simeq f_{VN}(0) \rho .
\end{eqnarray}
 Although this formula is valid at extremely low density,
 it is {\em useless} at nuclear matter density
 due to (\ref{nono}).

 Motivated by the formula (\ref{masss}), however,
 it is claimed 	in ref.\cite{Koike} that the mass shift is positive since
 $f_{VN}(0)$ is positive in a QCD sum rule estimate.
  It can be shown that  this claim is erroneous
 \cite{HLS}:

\noindent
\  \ \ \ (i) Use of the formula (\ref{masss})
 in nuclear matter is wrong from the outset.

\noindent
\ \ \ \ (ii) The calculation of $f_{VN}$ in \cite{Koike}
 is wrong due to the following reason.
 The author starts with the QCD  sum rules for the
 scattering amplitude $\Pi^{VN} = \langle N \mid {\rm T}J_{\mu}(x)
 J_{\nu} (0) \mid N \rangle $.
 For ${\rm Im} \Pi^{VN}$,
 \underline{three} phenomenological parameters are introduced
 (one of them is $f_{VN}(0)$). On the other hand,
  only \underline{two} sum rules can be obtained from the OPE of
 ${\rm Re} \Pi^{VN}$.
  Now,  one cannot solve three unknowns from two
 equations without using a magic.
   To get a valid estimate of $f_{VN}(0)$, one needs
 higher orders in OPE, which is practically very difficult to do
 for $\Pi^{VN}$.
 Thus it is impossible to get reliable
 result for the scattering length in QCD sum rules at the present stage.

\noindent
\  \ \ \ (iii)  The author also claims that,
 in the low density approximation (which is different from
 the fermi-gas approximation), (\ref{masss}) can be derived from the medium
 sum rules for $ \omega^2 \Pi^V(\omega)$. This statement is true, but
  it can be easily shown that the sum rules for $ \omega^2 \Pi^V(\omega)$
 have the same problem with (ii)   and are useless without
  the   information on  dim. 8 operators.
  Also, one cannot derive
  (\ref{masss}) from  $  \Pi^V(\omega)$ even in the low density approximation.
 Thus the above claim  does not have any influence on the validity
 of the sum rules used in the preceding sections.

  Let us summarize again the lessons we learned in this subsection: Firstly,
 the mass shift and the scattering length does not have
 direct connection in nuclear matter due to the momentum
 dependence of the $V-N$ forward scattering amplitude.
 Secondly, sum rules for the $V-N$ scattering amplitude cannot predict
 the $V-N$ scattering length without dimension 8 operators in OPE.
 Thirdly, sum rules for $\omega^2 \Pi^V(\omega)$
 does not work at all even in the vacuum without
 dimension 8 operators  and so does in the medium.
 Thus all the claims given in ref.\cite{Koike}
 are invalid. Also, only the consistent sum rules
 in medium currently available is the one starting
 from $\Pi^V(\omega)$ given in \cite{HL}.

There is a similar confusion on  the nucleon in nuclear matter \cite{KM},
 which is actually predated ref.\cite{Koike}.
 For the $N-N$ forward scattering amplitude $f_{NN}(p)$,
 one can prove that it has a huge $p$ dependence at low $p$
 using the
  low energy $N-N$ phase shift \cite{HLS}.
  The deutron resonance
 in $ ^3S_1$ channel and the strong attraction
 in $ ^1S_0$ channel near $p=0$ induce a rapid variation
 of $f_{NN}(p)$, which makes one impossible to
 approximate the amplitude by $f_{NN}(0)$ in the
 interval $0 < p < p_{_F}$.  Thus, the N-N scattering length and the
 optical potential for the nucleon in nuclear matter have no connection.

\subsection{Remarks on the effective theory approaches}

 There exist many attempts so far to calculate the
 spectral change of the vector mesons using effective  models.
 The calculation by Chin \cite{chin}
 using the Walecka model
 predict the increasing $\omega$-meson mass
 in medium due to the scattering process
\begin{eqnarray}
 \omega + N \rightarrow
 N \rightarrow \omega + N.
\end{eqnarray}
  For the $\rho$-meson, more sophisticated calculations
 including $\Delta$ and in-medium pion contributions
 predict a slight increase of the $\rho$-mass \cite{herman}.
  In all these calculations, only the effect
 of the polarization of the Fermi sea is taken into account.

  On the other hand, Kurasawa and Suzuki have stated in  clear terms
 that the mass of the $\omega$-meson is affected substantially
 by the vacuum polarization in medium
\begin{eqnarray}
 \omega \rightarrow N_* \bar{N}_* \rightarrow \omega,
\end{eqnarray}
 where  $N_*$ is the nucleon
  in nuclear medium \cite{kusu}. The vacuum polarization  dominates over the
 Fermi-sea polarization and leads decreasing
 vector meson mass. This conclusion was confirmed later
 by other authors \cite{others}
 and also generalized to the $\rho$-meson \cite{HS}.

 What is missing in the {\em Fermi sea} approaches \cite{chin,herman}
  is the
 effect of the scalar mean-field on the vector meson mass.
 On the other hand,  {\em Dirac sea} approaches \cite{kusu,others,HS}
  have close similarity with the other
 mean-field models
 such as those of Brown and Rho \cite{BR},
 Jaminon and Ripka \cite{JR}, and Saito and Thomas \cite{ST},
 which predict the decrease of the vector-meson masses.
   It is desirable to develop a unified  effective lagrangian
 which embodies the essential part of these approaches \cite{KH95}.

\section{Planned experiments -- What one should look for? --}

\subsection{finite $T$ case}
 As I have emphasized several times, the $\phi$ meson is an
 excellent candidate to see the dynamical critical phenomena
at finite $T$. Asakawa and Ko have proposed
 the so-called double $\phi$-peak signal which is shown
 in Fig.18 \cite{AK}.  Fig. 18(a) shows the time history
 of the hot region created by the relativistic heavy ion collisions.
 Central plateau in the figure shows a slow conversion from the
 quark gluon plasma phase to the hadronic phase.
 The first order phase transition is not necessary to have  this plateau:
 Rapid (but continuous) change of the entropy in a narrow
 temperature interval, which is actually seen on the lattice \cite{NC},
 is enough for the long duration of the plateau.
 $\phi$-mesons in the lepton pair spectrum should have two components if the
 initial temperature is high enough as Fig. 18 (b).
 One is the shifted $\phi$ decaying from the plateau region.
 The other is the $\phi$ decaying at later stages.
   Longer the duration time of the plateau, the
 higher the shifted peak-height grows. Such signals could be
 seen at RHIC and LHC where the quark-gluon plasma is expected to be
 created in the initial stage of the collisions.

\begin{figure}[t]
   \vspace{22pc}
  \caption{(a) Proper-time ($\tau$) history of the
 hot matter.  (b) Double $\phi$ peak expected in the lepton pair
 spectrum. The peak around 0.78GeV is the $\omega$ which
 is assumed not to have mass shift. The figures are taken from [42].}
\end{figure}

\subsection{finite $\rho$ case}

The vector meson mass shift 	at finite baryon density could be
 seen in heavy nuclei.
 There exit already two proposals to look for it \cite{SE}.
 One is by Shimizu et al.: They propose an experiment
 to  create $\rho$ and $\omega$
 in heavy nuclei using coherent photon - nucleus reaction and
  subsequently  detect the lepton pairs from $\rho$ and $\omega$.
 Enyo et al. propose to create $\phi$ meson in heavy nuclei
 using the proton-nucleus reaction and
 to measure kaon pairs as well as the lepton pairs.
 By doing this, one can study not only the mass shift
but also the change of the leptonic vs hadronic branching ratio
\begin{eqnarray}
r = \Gamma(\phi \rightarrow e^+ e^- ) /\Gamma (\phi \rightarrow K^+ K^- ),
\end{eqnarray}
which is sensitive to the change of the $\phi$-mass  as well as $K$-mass
 in medium.

In Table 3, some details about the above  planned experiments are summarized.
 There are also on-going heavy ion experiments at SPS (CERN) and AGS (BNL)
 where  high density matter is likely to be formed.
 In particular, CERES/NA45 at CERN recently reported an enhancement of the
 $e^+e^-$ pairs below the $\rho$ resonance, which is hard to be explained
 by the conventional sources of the lepton pairs \cite{ceres}.
 Also, E859 at BNL-AGS reported a possible spectral change of the
 $\phi$-peak in $K^+K^-$ spectrum \cite{ags}.
 If these effects are real, the shoulder structure of the spectrum
 expected by the mass shift of the vector mesons could be a
  possible explanation \cite{hatsuda95}.

\vspace{0.5cm}

\noindent
Table 3: Planned experiments aiming to detect the
 spectral change of vector mesons.

\vspace{8cm}

\newpage

\section{Concluding remarks}

The spectral change of the elementary excitations in medium
 is an exciting new possibility in QCD.
 By studying such phenomenon, one can learn the structure
 of the hadrons and the QCD ground state at finite $(T, \rho$)
 simultaneously.   Some theoretical models
 predict that the  light vector mesons ($\rho$, $\omega$ and $\phi$)
 are sensitive to the partial restoration of chiral symmetry
 in hot/dense medium.  These mesons are also experimentally good
 probes since they decay into lepton pairs which penetrate
 the hadronic medium without loosing much information.
  Thus, the lepton pair spectroscopy in QCD will tell us a
 lot about the detailed structure of the hot/dense matter, which
 is quite similar to the soft-mode spectroscopy
 by the photon and neutron scattering experiments in solid state physics.
 The theoretical approaches to study the spectral changes
 are still in the  primitive stage and  new methods beyond
 QCD sum rules and naive effective lagrangian approaches
 are called for.

\vspace{2cm}

\centerline{\bf{Acknowledgements}}

I thank the members of the organization committee
 for giving me an opportunity to give a talk at the
 9th Nishinomiya Yukawa-Memorial Symposium.
 I also thank M. Asakawa,  M. Doui, K. Kanaya,
 T. Kunihiro,   S. H. Lee and A. Ukawa for numerous discussions and
 useful information, and  J. Chiba,  H. Enyo,  K. Kurita, Y. Miake,
  H. Shimizu and K. Yagi for experimental information.
   I would like to dedicate this article to Prof. R. Tamagaki who
 has been giving me  continuous encouragement.

\end{document}